\documentclass[11pt]{article}

\usepackage[a4paper,margin=1in]{geometry}
\usepackage[T1]{fontenc}
\usepackage[utf8]{inputenc}
\usepackage{lmodern}
\usepackage{graphicx}
\usepackage{booktabs}
\usepackage{amsmath}
\usepackage{amssymb}
\usepackage{mathtools}
\usepackage{subcaption}
\usepackage{multirow}
\usepackage[table]{xcolor}
\usepackage{placeins}
\usepackage{makecell}
\usepackage{tabularx}
\usepackage{ragged2e}
\usepackage{wrapfig}
\usepackage{float}
\usepackage[accsupp]{axessibility}
\usepackage{hyperref}
\usepackage{microtype}

\begin{document}
\raggedbottom

\title{LITTA: Late-Interaction and Test-Time Alignment for Visually-Grounded Multimodal Retrieval}
\author{Seonok Kim\\Mazelone\\\texttt{seonokrkim@gmail.com}}
\date{}
\maketitle

\begin{abstract}
  Retrieving relevant evidence from visually rich documents such as textbooks, technical reports, and manuals is challenging due to long context, complex layouts, and weak lexical overlap between user questions and supporting pages. We propose LITTA, a query-expansion-centric retrieval framework for evidence page retrieval that improves multimodal document retrieval without retriever retraining. Given a user query, LITTA generates complementary query variants using a large language model and retrieves candidate pages for each variant using a frozen vision retriever with late-interaction scoring. Candidates from expanded queries are then aggregated through reciprocal rank fusion to improve evidence coverage and reduce sensitivity to any single phrasing. This simple test-time strategy significantly improves retrieval robustness while remaining compatible with existing multimodal embedding indices.

We evaluate LITTA on visually grounded document retrieval tasks across three domains: computer science, pharmaceuticals, and industrial manuals. Multi-query retrieval consistently improves top-$k$ accuracy, recall, and MRR compared to single-query retrieval, with particularly large gains in domains with high visual and semantic variability. Moreover, the accuracy--efficiency trade-off is directly controllable by the number of query variants, making LITTA practical for deployment under latency constraints. These results demonstrate that query expansion provides a simple yet effective mechanism for improving visually grounded multimodal retrieval.

\end{abstract}

\section{Introduction}
\label{sec:introduction}

Retrieving relevant evidence from visually rich documents remains a challenging problem in multimodal information retrieval. Documents such as textbooks, technical manuals, and scientific reports often contain complex layouts, diagrams, tables, and long contextual dependencies. As a result, evidence relevant to a user query may appear in visually structured regions of a page rather than in plain text passages. Recent multimodal retrieval systems address this challenge by encoding document pages with vision-language models and performing retrieval over page-level embeddings.

In this work, we emphasize evidence page retrieval: the retrieval target is not merely the correct document, but the specific page(s) that contain the supporting table, figure, or procedure. This page-level view is crucial for visually grounded downstream reasoning, because the decisive evidence is often localized and cannot be recovered from neighboring pages without retrieving it first.

Evidence page retrieval is difficult not only because of visual complexity, but also because the ``right'' page is frequently described in ways that are indirect from the user's perspective. A query may refer to an outcome (e.g., a failure mode), while the document uses part numbers and schematic labels; a question about a regulation may be answered by a table header rather than a paragraph; and a procedural question may be supported by a diagram plus a short caption. In these cases, retrieval must bridge a gap between the user-facing phrasing and the document's internal terminology and visual conventions.

Despite these advances, most retrieval systems rely on a single query representation to retrieve candidate pages. However, user questions often underspecify the relevant visual evidence, especially in domains with weak lexical overlap between the query and the document content. For example, a question about a technical concept may correspond to figures, diagrams, or descriptions that use different terminology or visual explanations. As a result, single-query retrieval may fail to retrieve relevant evidence pages even when the underlying retriever is strong.

Visually rich documents introduce failure modes that are not well captured by single-shot querying. First, user terminology often mismatches the phrasing used in manuals and regulatory reports, including abbreviations, part names, and compliance jargon. Second, evidence frequently appears in tables and figures, so the key semantics may be only weakly expressed in surrounding text. Third, a relevant page may support the answer implicitly through diagrams or parameter tables, leaving few lexical cues even when the page is decisive evidence.

Late-interaction multi-vector retrievers mitigate some of these issues by preserving fine-grained token--region interactions via ColBERT-style MaxSim scoring~\cite{khattab2020colbert}. Yet even strong retrievers can be brittle when constrained to a single query: the query may fail to express the right terminology or constraints needed to surface the evidence page.

In this work, we propose LITTA, a retrieval framework that improves multimodal document retrieval through query expansion. Given a user query, LITTA generates a small set of complementary query variants that preserve the original intent while emphasizing alternative terminology, constraints, and contextual cues. Each query variant retrieves candidate pages using a frozen late-interaction vision retriever, and we fuse the ranked lists across variants using Reciprocal Rank Fusion (RRF)~\cite{cormack2009rrf} to promote pages that are consistently ranked highly across variants. This test-time strategy increases the likelihood that at least one query variant aligns with the evidence page, improving coverage without retriever retraining. While multi-query retrieval increases online cost roughly linearly in the number of variants, small values (e.g., $Q{=}3$) provide a strong accuracy--efficiency trade-off.

Our pipeline is designed for practical deployment. Page embeddings are precomputed offline and stored in a compact index, while online computation scales primarily with the number of query variants and per-variant candidate depth. The retrieval cost is therefore controllable, enabling a favorable accuracy--efficiency trade-off in real-world settings.

Beyond improving raw retrieval metrics, we view query expansion as a robustness mechanism for visually grounded retrieval. When evidence is concentrated in visually dense regions (tables, plots, schematics), minor lexical shifts can substantially change which page regions are activated under late interaction. By issuing a small set of diverse but faithful variants, LITTA reduces sensitivity to any single phrasing and instead aggregates consistent retrieval signals across variants.

We evaluate LITTA on visually grounded document retrieval tasks across three domains: computer science, pharmaceuticals, and industrial manuals. Our experiments show that multi-query retrieval consistently improves top-$k$ accuracy, recall, and ranking metrics compared to single-query baselines. In particular, gains are larger in industrial manuals where terminology is heterogeneous and evidence is frequently embedded in figures and procedural tables, while improvements remain consistent in textbooks and pharmaceutical reports.

Our contributions are summarized as follows:
\begin{itemize}
\item We propose LITTA, a simple query-expansion framework for improving multimodal document retrieval without retriever retraining.
\item We introduce a multi-query retrieval strategy that pools candidates across query variants using late-interaction scoring and robust rank-level aggregation.
\item Extensive experiments across three visually rich document domains demonstrate consistent improvements over strong multimodal retrieval baselines, with particularly large gains in terminology- and layout-diverse manuals, and highlight practical accuracy--cost trade-offs.
\end{itemize}

\section{Related Work}
\label{sec:related_work}

\subsection{Visual Document Retrieval and Late Interaction}

Visually rich document retrieval requires grounding queries to page-level evidence that may be concentrated in localized regions such as tables, figures, or diagrams. Recent work has explored model- and system-level strategies for cross-modal retrieval and document understanding, spanning generative or coarse-to-fine retrieval formulations and unified visual--textual integration for document retrieval~\cite{fang-etal-2025-cart,Zhang_2025_CVPR,sun-etal-2025-unveil}. Parallel progress in long-document and multi-page understanding further highlights the need for robust evidence localization under complex layouts and long contexts~\cite{hu-etal-2025-mplug,ye2025mplugowl,zhu2025inteexpladva,deng2025longdocurl}.

Late-interaction multi-vector retrieval is particularly well-suited to this setting because it preserves token/region-level interactions rather than collapsing pages into a single embedding. Strong late-interaction embedding models for visual document retrieval, such as Nemotron ColEmbed V2~\cite{moreira2026nemotroncolembedv2topperforming}, enable efficient MaxSim-style matching and fine-grained evidence alignment. More broadly, structured retrieval and reasoning pipelines based on graphs and multi-agent systems have been studied for multimodal retrieval and grounding~\cite{bu-etal-2025-query,han2025retrievalaugmentedgenerationgraphsgraphrag,wang2025taming,lu2025karma,hong2025knowledgebasedvisualquestionanswer,lin-etal-2025-makar}. However, these approaches typically emphasize model-side advances or structured reasoning, whereas we focus on a retriever-agnostic, test-time strategy that improves page retrieval by diversifying queries and aggregating evidence across variants.

Recent work also shows that vision-language models can be used to construct efficient multi-vector embeddings directly in the visual space for document retrieval. For example, ColPali~\cite{faysse2025colpali} projects VLM patch features into a multi-vector representation and trains retrieval using ColBERT-style late interaction~\cite{khattab2020colbert}, enabling robust retrieval without brittle OCR or layout pipelines.

\subsection{Multimodal RAG Retrieval}

Multimodal RAG systems have been developed for visually rich documents and long-document settings, combining page retrieval with downstream multimodal generation~\cite{tanaka2025vdocrag,chen2025svrag,yu2025visrag,sun2025visrag20evidenceguidedmultiimage,shi2026urag,suri2024visdommultidocumentqavisually,Cho2024M3DocRAG,wu2025molorag}. In parallel, benchmarks and analyses evaluate retrieval-augmented multimodal systems across domains and tasks, including large-scale user-centric evaluations and domain-specific resources~\cite{li2024benchmarkingmultimodalretrievalaugmented,hu2024mragbench,chan2025mmuragent,zhao2025finragbench,xia2024mmedrag}.

Despite rapid progress, reliable evidence page retrieval remains a bottleneck when answers are distributed across multiple pages and the query--document lexical overlap is weak. Our work targets this bottleneck directly: we improve evidence page retrieval at test time by generating complementary query variants and aggregating their retrieval signals, without retriever retraining or architectural changes.

\subsection{Query Optimization and Retrieval-Time Adaptation}

Retrieval-time adaptation and query optimization aim to improve evidence retrieval without retraining the underlying retriever, often by using agents, feedback signals, or complementary modules to refine queries at inference time~\cite{wang2025vidorag,liu2025cadencerag,biswas2025raven,blau2024gramglobalreasoningmultipage,han2025mdocagent,su2025skvqasynthet,wangretrieval,uzan2026guided}. These methods demonstrate that test-time optimization can increase robustness under distribution shifts and complex multimodal inputs.

Within this broader space, a practical line of work improves retrieval by issuing multiple complementary queries and aggregating their retrieval signals. Multi-query strategies are attractive for visually grounded settings because the same evidence page can be surfaced by different cues: a part name vs.\ a schematic label, a high-level concept vs.\ a table header, or an acronym vs.\ its expanded form. Crucially, these gains can often be realized without changing the retriever, by shifting the adaptation effort to the query side.

Our approach follows the same test-time spirit but focuses on a simple and scalable mechanism: multi-query generation that produces semantically diverse query variants and improves coverage under terminology mismatch, combined with late-interaction scoring for fine-grained evidence localization. To aggregate evidence across variants, we use Reciprocal Rank Fusion (RRF)~\cite{cormack2009rrf}, a robust rank-level fusion method that emphasizes candidates that consistently appear near the top across lists. Unlike approaches that require additional reasoning modules or iterative optimization loops, LITTA provides a lightweight retrieval-time improvement that is compatible with existing embedding indices.

\section{Method}
\label{sec:method}

\begin{figure}[t]
\centering
\includegraphics[width=\linewidth]{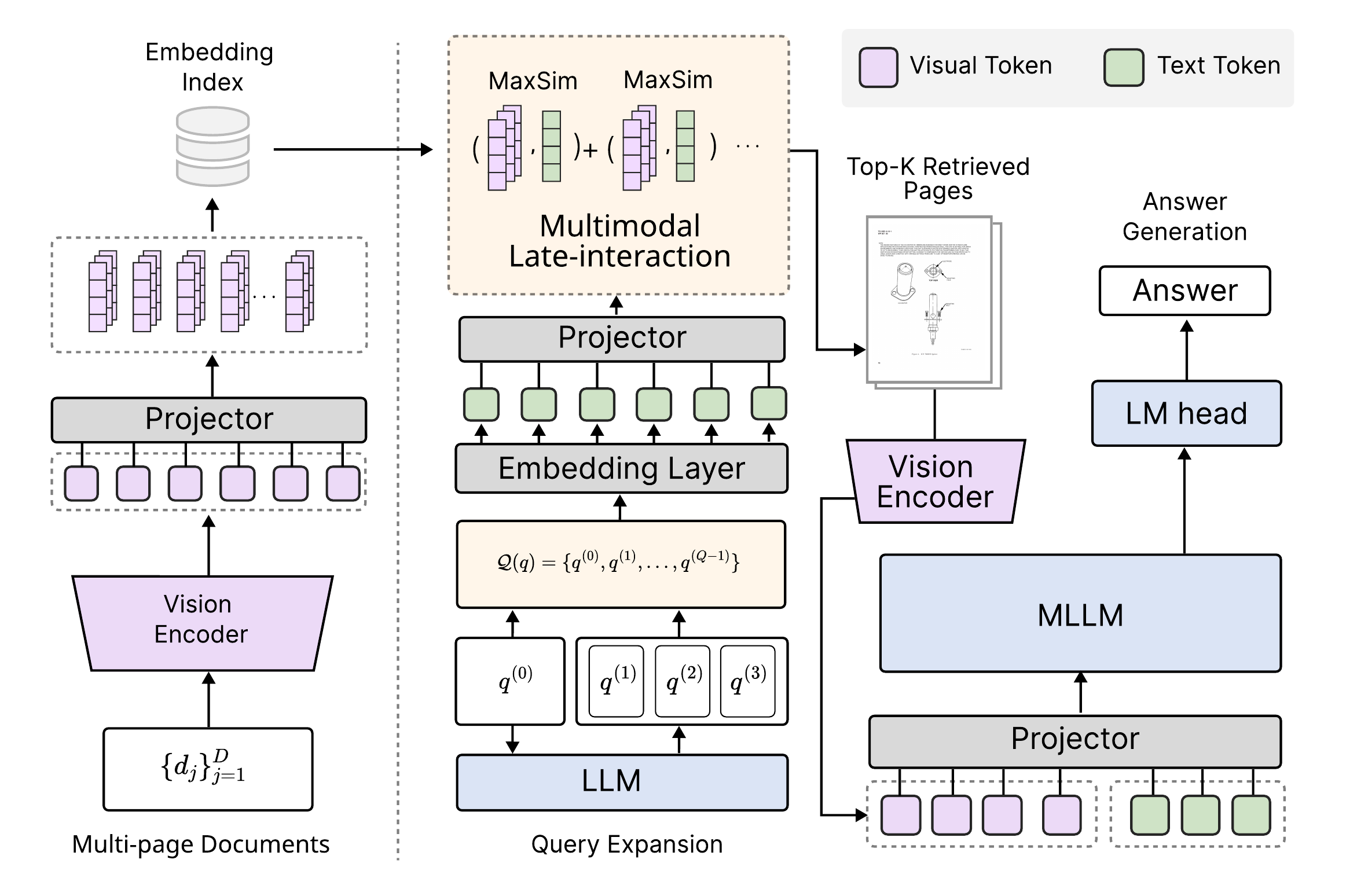}
\caption{Overview of LITTA. Left: we pre-compute and index page embeddings for a multi-page document collection. Middle: a query is expanded into complementary variants. Right: each expanded query retrieves top-$k$ candidate pages via late-interaction scoring; we aggregate candidates across expansions using reciprocal rank fusion, optionally re-rank a small subset, and pass top-$K$ evidence pages to the generator.}
\label{fig:framework}
\end{figure}

We introduce LITTA (Late-Interaction Test-Time Alignment), a lightweight framework for improving retrieval from visually rich document collections through query expansion and multi-query evidence aggregation. Figure~\ref{fig:framework} illustrates the overall pipeline. Given a user query, LITTA first generates multiple complementary query variants using a large language model (LLM). Each expanded query is then encoded and used to retrieve candidate document pages from a pre-computed multimodal embedding index via late-interaction scoring. Finally, candidate pages retrieved from different query variants are aggregated to produce a unified ranked list of evidence pages for downstream answer generation.

\subsection{Problem Formulation}

We consider the problem of retrieving relevant pages from a collection of visually rich multi-page documents. Let $D=\{p_i\}_{i=1}^{N}$ denote a page collection, where each page may contain a mixture of text, diagrams, tables, and other visual structures. Given a user query $q$, the goal is to retrieve the top-$K$ pages that contain evidence relevant to the query.

Recent multimodal retrieval systems represent each page as a set of visual embeddings extracted from a vision-language encoder. However, traditional retrieval encodes the query once and performs retrieval with a single query representation. This can fail when the lexical form of the query does not directly match the terminology or visual explanations present in the document pages. LITTA addresses this limitation by expanding the query into multiple complementary variants before retrieval.

Importantly, we do not assume that the correct evidence can be isolated into short passages or that OCR reliably captures the salient information in tables and figures. Instead, we retrieve whole pages and rely on late interaction to match query tokens against localized visual regions. Under this page-level protocol, the primary risk of single-query retrieval is under-specification: the query may omit the attribute, synonym, or domain-specific label that is necessary to activate the correct region on the evidence page.

\subsection{Document Embedding and Indexing}

Each page is encoded by a frozen vision encoder followed by a lightweight projection layer, producing a multi-vector representation:
\begin{equation}
E(p_i)=\left\{\mathbf{v}_i^{(1)}, \mathbf{v}_i^{(2)}, \dots, \mathbf{v}_i^{(M)}\right\}, \quad \mathbf{v}_i^{(m)}\in\mathbb{R}^{d}.
\end{equation}
Unlike single-vector retrieval, storing multiple visual token embeddings preserves fine-grained visual and layout information (e.g., table cells and diagram regions). In practice, page images are processed into visual token sequences (potentially via tiling for higher visual granularity), and the projection layer maps these tokens into the retrieval embedding space. All page embeddings are pre-computed offline and stored in an embedding index for efficient retrieval (Figure~\ref{fig:framework}, left). In our implementation, we use Nemotron ColEmbed V2 as the frozen late-interaction vision retriever/encoder~\cite{moreira2026nemotroncolembedv2topperforming}.

This separation between offline indexing and online querying is central to LITTA's design: the index is constructed once, while query-side adaptation can be changed at test time without re-encoding the corpus. As a result, the cost of multi-query retrieval is dominated by the number of query variants and the per-variant candidate depth, rather than by any additional corpus processing.
Concretely, the index supports approximate nearest neighbor lookup to produce a short candidate list for each query variant, after which late interaction is applied only to the shortlist. The number of stored visual tokens $M$ depends on the page resolution and any tiling strategy; we keep the page encoding procedure fixed across runs so that improvements are attributable to query-side adaptation.

\subsection{Query Expansion with Large Language Models}

The first component of LITTA generates a set of expanded queries that capture different semantic interpretations of the original question. Given an input query $q$, an LLM produces a small set of expanded queries, yielding $Q$ total query variants:
\begin{equation}
\mathcal{Q}(q)=\left\{q^{(0)}, q^{(1)}, \dots, q^{(Q-1)}\right\}, \quad q^{(0)}=q,
\end{equation}
where $q^{(0)}$ corresponds to the original query and the remaining $Q-1$ queries represent alternative formulations generated by the language model.

These expanded queries aim to capture diverse lexical and semantic expressions that may correspond to the same underlying information need. For example, a question about a technical component might correspond to terminology appearing in diagrams, figure captions, or descriptive paragraphs using different wording. By generating multiple complementary queries, LITTA increases the likelihood that at least one query variant aligns well with the textual or visual representation of relevant document pages (Figure~\ref{fig:framework}, middle).

In our implementation, query variants are generated by a lightweight LLM~\cite{gemma_3n_2025}:
\begin{equation}
q^{(i)} = f_{\text{LLM}}^{(i)}(q), \quad i=1,\dots,Q-1.
\end{equation}
The generation objective is to preserve the original intent while encouraging diversity across variants, for instance by injecting likely synonyms, abbreviations, or domain-specific names that may appear in figure captions or schematic labels. If LLM-based generation is unavailable, we fall back to a deterministic query-derived expansion so that retrieval remains stable and comparable across runs. In practice, we keep $Q$ small (e.g., $Q{=}3$ total queries) to balance coverage gains against the additional online retrieval cost.

Each query is encoded into a sequence of token embeddings:
\begin{equation}
E\!\left(q^{(i)}\right)=\left\{\mathbf{t}^{(i)}_1,\dots,\mathbf{t}^{(i)}_{L_i}\right\}.
\end{equation}

\subsection{Late-Interaction Retrieval}
\label{sec:method_late_interaction}

After generating expanded queries, each query variant is independently used to retrieve candidate pages from the document index. Following recent multimodal retrieval approaches, we represent each document page using a set of visual token embeddings extracted from a vision encoder, and represent each query using a set of token embeddings. Retrieval is performed using a late-interaction scoring mechanism, where the similarity between a query and a document page is computed by aggregating token-level similarities.

Specifically, we use a MaxSim scoring function that computes the maximum similarity between each query token and the set of page embeddings. The final similarity score between query $q^{(i)}$ and page $p_j$ is obtained by summing the maximum similarities across all query tokens:
\begin{equation}
\text{Score}\!\left(q^{(i)}, p_j\right)=\sum_{\ell=1}^{L_i}\max_{m\in\{1,\dots,M_j\}}
\left(\mathbf{t}^{(i)}_{\ell}\right)^\top \mathbf{v}^{(m)}_{j}.
\end{equation}
This late-interaction formulation preserves fine-grained interactions between query tokens and visual tokens, allowing the retriever to capture detailed semantic correspondences between the query and document content (e.g., tables, diagrams, and captions). In practice, we first use the embedding index to identify a shortlist of candidates per variant and then apply late interaction to score only this shortlist. For each query variant, we retain the top-$k$ candidate pages based on the late-interaction similarity score (Figure~\ref{fig:framework}, right).
For scalability, scoring is performed in batches and we keep a fixed per-variant candidate depth $k$ before aggregation.

\subsection{Candidate Aggregation Across Expanded Queries}

Since each expanded query retrieves its own ranked list of candidate pages, LITTA aggregates candidates across query variants to produce a unified ranking. Let $\mathcal{C}_i$ denote the set of top-$k$ pages retrieved for query variant $q^{(i)}$. The final candidate pool is constructed by combining all retrieved pages:
\begin{equation}
\mathcal{C}=\bigcup_{i=0}^{Q-1}\mathcal{C}_i.
\end{equation}

We fuse the ranked lists using Reciprocal Rank Fusion (RRF). Let $r_i(p)$ denote the rank position of page $p$ under query variant $q^{(i)}$. The fused score is:
\begin{equation}
\mathrm{RRF}(p)=\sum_{i=0}^{Q-1}\frac{1}{k_{\mathrm{RRF}} + r_i(p)}.
\end{equation}
Here $\mathcal{C}_i$ contains the top-$k$ candidates returned by variant $q^{(i)}$, and $r_i(p)$ is the 1-indexed rank of page $p$ within that list (we treat $r_i(p)=\infty$ if $p\notin\mathcal{C}_i$ so the corresponding term contributes approximately zero). RRF~\cite{cormack2009rrf} is a rank-level aggregation method that is robust to score scale differences across query variants and emphasizes pages that are consistently ranked highly. The constant $k_{\mathrm{RRF}}$ controls how quickly the contribution decays with rank; we use the standard setting ($k_{\mathrm{RRF}}{=}60$) unless stated otherwise.
The aggregated candidate set is then sorted according to $\mathrm{RRF}(p)$, and the top-$K$ pages are selected as the final retrieval output. Optionally, a lightweight re-ranking stage can be applied to a small subset of the top fused candidates to refine the final ordering before passing selected evidence pages to the answer generation module.

\subsection{Answer Generation}

The selected evidence pages are passed to a multimodal large language model (MLLM) for answer generation. In our implementation, we use Qwen3-VL-4B-Instruct~\cite{bai2025qwen3vltechnicalreport} to generate the final answer conditioned on the query and the top-$K$ retrieved evidence pages $\{p_{1},\dots,p_{K}\}$. Visual and textual representations are projected into a shared embedding space and processed by the language model head:
\begin{equation}
\hat{y} = g_{\phi}(q, \{p_{1},\dots,p_{K}\}).
\end{equation}
Compared to single-query retrieval, LITTA improves recall by leveraging complementary query variants while preserving fine-grained multimodal interactions through late-interaction scoring.

\subsection{Complexity Analysis}

Let $M$ be the number of visual tokens per page, and let $L$ be the number of query tokens. Late-interaction scoring for one query variant against $k$ candidate pages has complexity:
\begin{equation}
\mathcal{O}(k \cdot L \cdot M).
\end{equation}
For $Q$ query variants, the total complexity becomes:
\begin{equation}
\mathcal{O}(Q \cdot k \cdot L \cdot M).
\end{equation}
In practice, $k$ is the per-variant candidate depth returned by approximate nearest neighbor search over the embedding index, and late-interaction scoring is applied only to this shortlist. With pre-computed page embeddings, the online cost scales primarily with $Q$ and $k$, making the accuracy--efficiency trade-off directly controllable by the number of query variants and candidate depth.

\section{Experiments and Results}
\label{sec:experimentation}

We evaluate how multi-query expansion improves page-level retrieval on visually rich documents. Our focus is on evidence page retrieval, where the retriever must identify the correct page(s) that contain supporting tables, figures, and text for a query.

\subsection{Experimental Setup}
\label{sec:exp_setup}

\definecolor{oursbg}{RGB}{255,250,235}
\begin{table}[t]
\centering
\small
\renewcommand{\arraystretch}{1.15}
\setlength{\tabcolsep}{5pt}
\begin{tabular}{l c c c c c}
\toprule
\textbf{Model} & \textbf{Size} & \textbf{C.S.} & \textbf{Phar.} & \textbf{Ind.} & \textbf{Avg.} \\
\midrule
\multicolumn{6}{l}{\textcolor{gray}{\textit{Textual Retrievers}}} \\
BGE-M3 & 0.6B & 58.0 & 52.0 & 28.5 & 46.2 \\
Jina-v4 (Textual) & 3B & 64.3 & 54.9 & 38.4 & 52.5 \\
Qwen3 & 8B & 71.7 & 59.2 & 40.4 & 57.1 \\
Qwen3.5 & 8B & 73.7 & 67.0 & 42.6 & 61.1 \\
\midrule
\multicolumn{6}{l}{\textcolor{gray}{\textit{Visual Retrievers}}} \\
Nomic & 7B & 66.6 & 58.9 & 37.9 & 54.5 \\
ColQwen2.5 & 3B & 72.3 & 57.9 & 41.3 & 57.2 \\
ColEmbed & 1B & 71.3 & 62.6 & 46.6 & 60.2 \\
ColEmbed & 3B & 75.2 & 63.7 & 47.1 & 62.0 \\
ColNomic & 3B & 72.7 & 61.1 & 47.4 & 60.4 \\
ColNomic & 7B & 76.2 & 62.3 & 50.1 & 62.9 \\
Jina-v4 (Visual) & 3B & 71.8 & 63.1 & 50.4 & 61.8 \\
ColEmbed-v2 & 3B & 77.1 & 66.0 & 51.7 & 64.9 \\
\addlinespace[1mm]
\rowcolor{oursbg}
LITTA (q=1) & 3B & 79.3 & 74.3 & 39.05 & 64.2 \\
\rowcolor{oursbg}
LITTA (q=3) & 3B & \textbf{80.2} & \textbf{74.5} & 53.67 & 69.5 \\
\rowcolor{oursbg}
LITTA (q=5) & 3B & 80.2 & 74.3 & \textbf{61.39} & \textbf{72.0} \\
\bottomrule
\end{tabular}
\vspace{0.8em}
\caption{Retrieval performance (NDCG@10) across three domains. Multi-query retrieval with query variants ($q>1$) consistently improves ranking quality over single-query retrieval ($q=1$), with the largest gains in Industrial manuals where terminology and layout variability reduce lexical overlap. \textbf{Bold} indicates best.}
\label{tab:tb_retrieval_performance}
\end{table}

\paragraph{Datasets.}
We evaluate on three visually grounded domains from the ViDoRe V3 benchmark~\cite{loison2026vidorev3comprehensiveevaluation}:
Computer Science Textbooks (C.S.), Pharmaceutical Reports (Phar.), and Industrial Maintenance Manuals (Ind.). Each document is split into pages and retrieval is performed at the page level. Dataset statistics are summarized in Appendix~\ref{tab:appendix_1_tb_dataset_statistics}.
\looseness=-1
Each query is associated with one or more evidence pages, reflecting that supporting information may span a table, a figure and its caption, or multiple consecutive pages. We evaluate retrieval over full pages to capture such visually localized evidence without assuming that relevant content can be isolated into short text passages.
This setup emphasizes robustness to layout and terminology: the gold evidence page may contain the decisive signal in a schematic, a parameter table, or a short caption, while the query may describe the same concept using user-facing phrasing.
\paragraph{Vision retriever (page-image).}
We use a frozen late-interaction vision retriever (Nemotron ColEmbed V2)~\cite{moreira2026nemotroncolembedv2topperforming}. Given a text query, it produces a multi-vector query embedding; given a page image, it produces a multi-vector page embedding, and similarity is computed via ColBERT-style MaxSim~\cite{khattab2020colbert}. We precompute page embeddings for each corpus and store them for efficient retrieval.

\paragraph{Multi-query expansion ($Q$).}
We compare query settings with expansion counts $Q \in \{1,3,5\}$, where $Q=1$ corresponds to a single original query and $Q>1$ corresponds to multi-query retrieval.
For multi-query, we generate $Q-1$ expanded queries (query variants) from the original query using a query-expansion generator. In our experiments, we use the Gemma 3N e4b variant as the expansion model when enabled~\cite{gemma_3n_2025}. Each query variant retrieves a ranked list of candidate pages, and we fuse these lists using Reciprocal Rank Fusion (RRF) to emphasize pages that are consistently ranked highly across variants.
We use RRF as a rank-level fusion method because it is stable under heterogeneous query variants and does not require calibrating similarity scores across variants~\cite{cormack2009rrf}.
Unless stated otherwise, we use $k_{\mathrm{RRF}}=60$, retrieve 20 candidates per query variant, and report the final top-20 pages after fusion.

\paragraph{Implementation details.}
Our pipeline is strictly retrieval-side: the vision retriever remains frozen and all improvements come from test-time query optimization. In practice, query variants are lightweight reformulations that target terminology mismatch (e.g., abbreviations, part names, or domain-specific phrasing) while keeping the original intent intact. To ensure stable evaluation, when query expansion is unavailable for a query (e.g., transient generation failure), we fall back to a deterministic query-derived expansion so that retrieval can proceed without interrupting the run.
Page embeddings are computed offline once per corpus and stored in an index, so online computation is dominated by (i) encoding each query variant and (ii) scoring a fixed candidate depth per variant before fusion. This makes the latency--accuracy trade-off transparent: increasing $Q$ increases cost roughly linearly, while fusion and final ranking are inexpensive relative to retrieval.

\paragraph{Baselines.}
We compare against strong textual and visual retrievers reported in Table~\ref{tab:tb_retrieval_performance}. Textual retrievers operate on page text, while visual retrievers encode page images and perform late-interaction matching. We use ColEmbed-v2 as a representative strong late-interaction vision retriever baseline and report both domain-wise and average performance.
In the Visual Retrievers block of Table~\ref{tab:tb_retrieval_performance}, Nomic refers to Nomic Embed~\cite{nussbaum2024nomic}.
For LITTA, we keep the underlying retriever fixed and vary only the query-side configuration ($Q$) and fusion strategy, isolating the contribution of multi-query retrieval.
This evaluation protocol is intentionally conservative: improvements reflect query-side coverage gains rather than architectural changes or retriever retraining.

\paragraph{Evaluation metrics.}
Each run logs a ranked list of retrieved page IDs and a set of relevant (gold) page IDs for each query. We report complementary page-level retrieval metrics averaged over test queries.
Top-$k$ accuracy (Hit@$k$) measures whether at least one relevant page appears in the first $k$ retrieved pages (we report curves for $k\in\{1,\ldots,10\}$).
Recall@$k$ measures evidence coverage, i.e., $\lvert \text{top-}k \cap \text{relevant}\rvert / \lvert \text{relevant}\rvert$ (we report Recall@5 and Recall@10).
MRR@$k$ measures how early the first relevant page is retrieved, i.e., $1/\text{rank}$ of the first relevant item within the top-$k$ (or $0$ if none is retrieved).
We additionally report NDCG@10 as the primary summary in Table~\ref{tab:tb_retrieval_performance}, capturing overall ranking quality while emphasizing early positions. Unless stated otherwise, values in tables are reported as percentages.

\subsection{Query Expansion Improves Retrieval}
\label{sec:exp_retrieval}

\begin{figure}
    \centering
    \includegraphics[width=\linewidth]{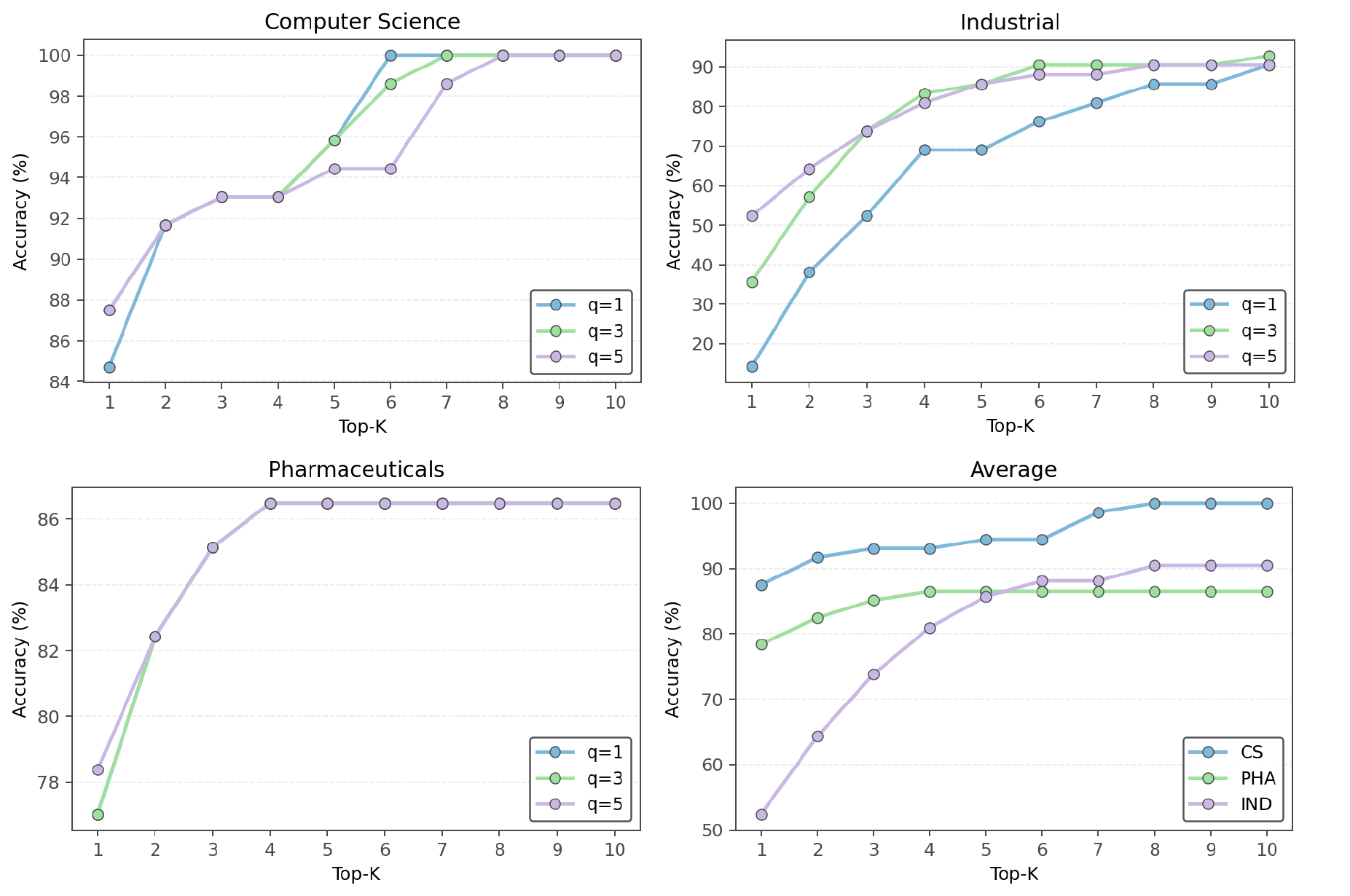}
    \caption{Top-$k$ accuracy (Hit@$k$) curves for multi-query settings $Q\in\{1,3,5\}$ across domains and the average.}
    \label{fig:graph}
    \end{figure}

We first measure retrieval performance as the number of expansion queries increases.
Table~\ref{tab:tb_retrieval_performance} shows that multi-query retrieval improves domain-wise retrieval quality over single-query retrieval. Gains are strongest in industrial manuals, where terminology and layout variation can cause single-query retrieval to miss relevant evidence pages. In contrast, C.S.\ textbooks are more lexically consistent, yielding smaller but still positive improvements.
For LITTA, we keep the underlying retriever fixed and vary only the query-side configuration ($Q$) and fusion strategy, isolating the contribution of multi-query retrieval.
Figure~\ref{fig:graph} reports Hit@$k$ curves. Increasing $Q$ consistently improves early retrieval, especially for Industrial manuals, indicating that query variants retrieve complementary evidence pages that are missed by the original query.
We also observe diminishing returns beyond a small number of variants: moving from $Q{=}1$ to $Q{=}3$ yields the most consistent improvements, while $Q{=}5$ can further improve difficult cases at a higher cost and with occasional noise from less faithful variants.

\subsection{Retrieval Recall and Multi-Query Expansion}
\label{sec:exp_recall}

We evaluate recall at top-5 and top-10 under multi-query retrieval.

\begin{table}[t]
\centering
\small
\renewcommand{\arraystretch}{1.15}
\setlength{\tabcolsep}{5pt}
\begin{tabular}{l c c c c c c c c}
\toprule
& \multicolumn{2}{c}{\textbf{C.S.}} & \multicolumn{2}{c}{\textbf{Phar.}} & \multicolumn{2}{c}{\textbf{Ind.}} & \multicolumn{2}{c}{\textbf{Avg.}} \\
\cmidrule(lr){2-3} \cmidrule(lr){4-5} \cmidrule(lr){6-7} \cmidrule(lr){8-9}
\textbf{Methods} & R@5 & R@10 & R@5 & R@10 & R@5 & R@10 & R@5 & R@10 \\
\midrule
ColEmbed-v2 & 67.07 & 82.80 & 72.13 & 77.24 & 59.81 & 73.96 & 66.34 & 78.00 \\
LITTA ($q$=1) & 67.95 & 82.64 & 72.11 & 77.74 & 43.89 & 65.52 & 61.32 & 75.30 \\
LITTA ($q$=3) & \textbf{68.22} & \textbf{83.32} & \textbf{73.33} & 77.27 & 57.63 & \textbf{73.76} & \textbf{66.39} & \textbf{78.45} \\
LITTA ($q$=5) & 65.74 & 83.11 & 72.65 & \textbf{77.74} & \textbf{60.37} & \textbf{73.76} & 66.25 & 78.01 \\
\bottomrule
\end{tabular}
\vspace{0.8em}
\caption{Recall@5 and Recall@10 (\%). ColEmbed-v2: base retriever; LITTA ($q$=$k$): multi-query retrieval with $k$ expanded queries. Bold indicates the best value per column; ties are bolded.}
\label{tab:tb_retrieval_performance_recall}
\end{table}

Table~\ref{tab:tb_retrieval_performance_recall} reports Recall@5 and Recall@10 for $Q \in \{1,3,5\}$. Increasing query count generally improves evidence coverage, indicating that expansions help retrieve additional relevant pages that may contain tables, figures, or alternative phrasing not captured by the original query.
These improvements are most pronounced in the Industrial domain, where visually grounded evidence is often described with domain-specific terminology and abbreviated labels, making it harder for a single query to match the relevant page.
Notably, recall gains indicate that query variants often recover \emph{additional} relevant pages rather than simply reordering the same set, which is important when downstream reasoning benefits from multiple pieces of evidence (e.g., a schematic plus an accompanying parameter table).

\subsection{Efficiency and Qualitative Analysis}
\label{sec:exp_efficiency}

\paragraph{Computational efficiency.}
Multi-query expansion introduces inference overhead proportional to the number of sub-queries because retrieval is executed $Q$ times. However, the observed gains in Recall and MRR make the approach practical for evidence-heavy domains. Across our experiments, $Q=3$ provides the best accuracy--cost trade-off, yielding consistent improvements with moderate additional retrieval cost.
\looseness=-1
From a system perspective, the multi-query overhead is predictable and can be tuned: practitioners can cap candidate depth per variant and choose a small $Q$ to meet latency budgets, while still benefiting from the robustness of fusion across variants.

\paragraph{Qualitative analysis.}
Single-query retrieval often misses relevant pages when user terminology differs from document phrasing (e.g., abbreviations, part names, or regulatory jargon) or when the evidence is primarily in tables/figures. Expansion queries recover complementary evidence pages that single-query retrieval fails to retrieve, improving evidence coverage and downstream answer support quality.
We observe that query variants are especially helpful when the relevant page uses alternative naming conventions (e.g., part numbers or abbreviations) or when evidence is concentrated in visually dense regions such as tables, where small lexical changes can substantially alter matching behavior under late interaction.
In contrast, when the corpus is lexically consistent (e.g., pedagogical textbooks), the original query already aligns well with page content, and multi-query retrieval mainly provides modest improvements by surfacing pages that express the same concept with slightly different phrasing.

\section{Conclusion}
\label{sec:conclusion}

We presented LITTA, a simple retrieval framework for evidence page retrieval in visually rich documents. LITTA improves multimodal retrieval through query-side test-time optimization: it generates a small set of complementary query variants and aggregates their retrieval signals using a frozen late-interaction vision retriever and Reciprocal Rank Fusion. Because it operates purely at retrieval time, LITTA is compatible with existing multimodal embedding indices and does not require retriever retraining or architectural modifications.

Across three visually grounded domains (computer science textbooks, pharmaceutical reports, and industrial manuals), multi-query retrieval consistently improves top-$k$ accuracy, recall, and ranking quality over single-query baselines. Gains are especially large for industrial manuals, where evidence is frequently encoded in schematics and procedural tables and document terminology is heterogeneous. These findings highlight that query diversity is a practical lever for improving evidence localization, even when the underlying retriever is already strong.

LITTA also has limitations. Multi-query retrieval increases online cost roughly linearly in the number of variants, and the quality of expansion affects the precision of retrieved candidates. Future work will explore adaptive strategies that select the number and type of variants based on query difficulty, incorporate lightweight re-ranking to improve early precision, and extend retrieval-time optimization to settings with richer feedback signals from downstream reasoning. Overall, our results suggest that query expansion is a lightweight yet effective mechanism for strengthening visually grounded multimodal retrieval in real-world document collections.

\bibliographystyle{splncs04}
\bibliography{main}

\appendix

%%%%%%%%%%%%%%%%%%%%%%%%%%%%%%%%%%%%%%%%%%%%%%%%%%%%%%%%%%%%%
\section{Dataset Statistics and Domain Characteristics}
\label{sec:appendix_dataset}
%%%%%%%%%%%%%%%%%%%%%%%%%%%%%%%%%%%%%%%%%%%%%%%%%%%%%%%%%%%%%

We evaluate LITTA on visually rich, multi-page document collections spanning education, pharmaceutical regulation, and industrial maintenance domains. Retrieval is performed at the page level because relevant evidence is often localized to a specific table, figure, or procedure page.

\begin{itemize}
\item Computer Science Textbooks (C.S.) contain structured educational material with dense textual explanations and diagrams.
\item FDA Reports (Phar.) consist of regulatory presentations and scientific documents with frequent tables, charts, and numerical evidence.
\item USAF Technical Orders (Ind.) include industrial maintenance manuals characterized by procedural steps and schematics.
\end{itemize}

Table~\ref{tab:appendix_1_tb_dataset_statistics} reports dataset statistics for the three evaluation corpora, including the size of the page collection and the number of evaluation queries. These statistics contextualize the relative difficulty of each domain (e.g., corpus scale and layout diversity) and motivate our page-level evaluation protocol, where success depends on retrieving the specific page(s) that contain the relevant evidence.

\looseness=-1
These domains stress different aspects of visually grounded retrieval. Textbooks tend to use consistent pedagogical terminology and provide longer surrounding explanations, while pharmaceutical reports frequently compress key evidence into tables and plots with domain-specific abbreviations. Industrial manuals further introduce heterogeneous naming conventions (part numbers, acronyms, and schematic labels) and highly structured layouts, where evidence may be concentrated in a diagram region with minimal nearby text. This diversity makes evidence page retrieval a natural testbed for query-side robustness mechanisms such as multi-query expansion.

%%%%%%%%%%%%%%%%%%%%%%%%%%%%%%%%%%%%%%%%%%%%%%%%%%%%%%%%%%%%%
\section{Prompts Used in LITTA}
\label{sec:appendix_prompts}
%%%%%%%%%%%%%%%%%%%%%%%%%%%%%%%%%%%%%%%%%%%%%%%%%%%%%%%%%%%%%

\begin{figure}[t]
\centering
\includegraphics[width=\linewidth]{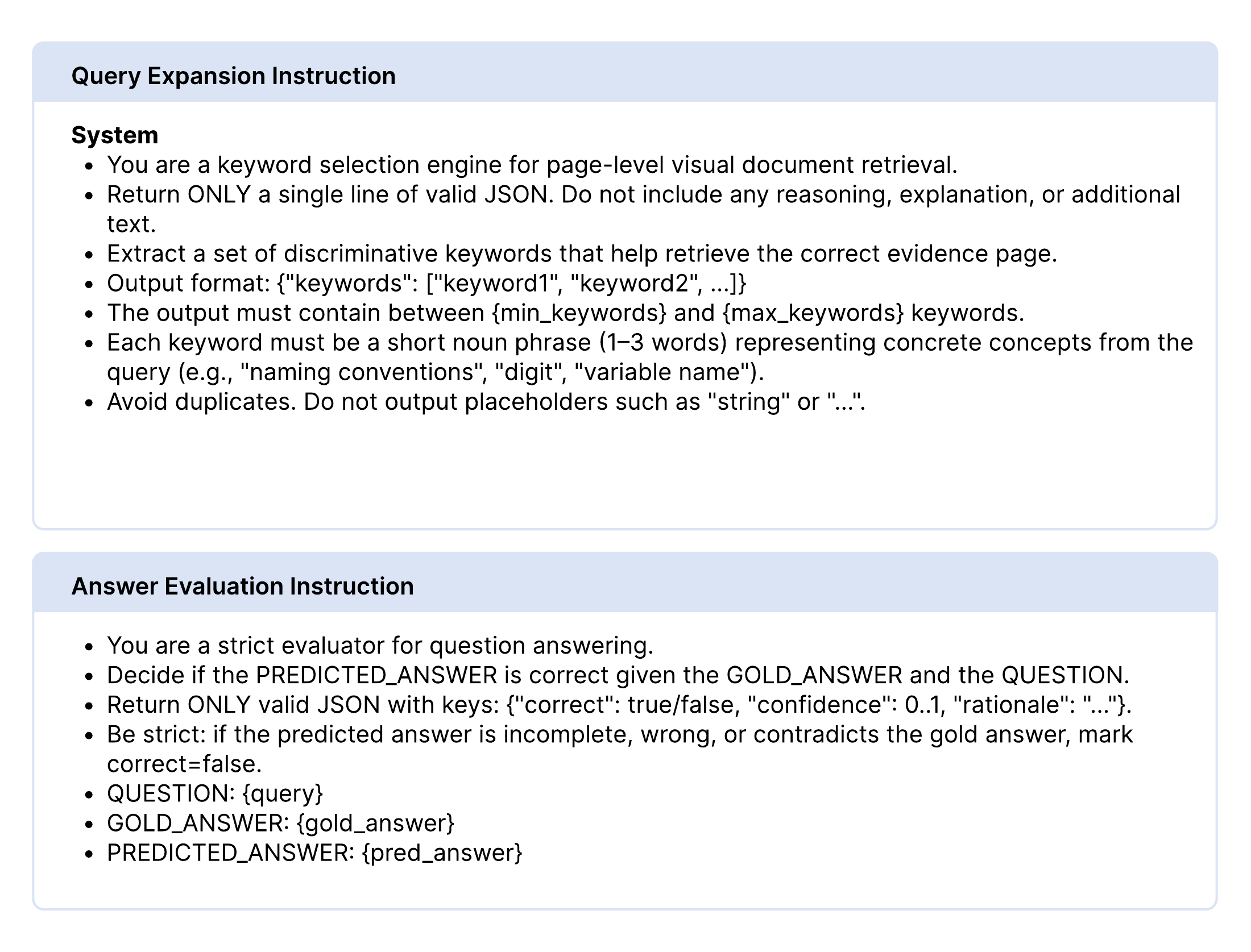}
\caption{Prompt templates used for query expansion (query variant generation) and answer evaluation.}
\label{fig:appendix_instruction}
\end{figure}

We provide the prompt templates used in our pipeline. Figure~\ref{fig:appendix_instruction} contains (i) the query-expansion prompt used to generate query variants and (ii) the answer evaluation prompt used for strict correctness checking in auxiliary experiments.
\begin{table}[t]
\small
\centering
\resizebox{\linewidth}{!}{%
\setlength{\tabcolsep}{4pt}
\renewcommand{\arraystretch}{1.15}
\begin{tabular}{l l c c c l}
\toprule
\textbf{Corpus} &
\textbf{Domain(s)} &
\textbf{\# Docs} &
\textbf{\# Pages} &
\textbf{\# Queries$^{*}$} &
\textbf{Main modalities} \\
\midrule

Computer Science Textbooks &
Computer Science / Education &
2 & 1360 & 215 &
Books \\

FDA Reports &
Pharmaceuticals &
52 & 2313 & 364 &
Slides, Books \\

USAF Technical Orders &
Industrial Maintenance &
27 & 5244 & 283 &
Manuals \\
\bottomrule
\end{tabular}%
}

\vspace{0.9em}
\caption{Dataset statistics for the three evaluation corpora used in our experiments.}
\label{tab:appendix_1_tb_dataset_statistics}
\end{table}

%%%%%%%%%%%%%%%%%%%%%%%%%%%%%%%%%%%%%%%%%%%%%%%%%%%%%%%%%%%%%
\section{Implementation Details}
\label{sec:appendix_impl}
%%%%%%%%%%%%%%%%%%%%%%%%%%%%%%%%%%%%%%%%%%%%%%%%%%%%%%%%%%%%%

\subsection{Pre-computed Page Embeddings}

We encode each page image into a multi-vector representation using a frozen late-interaction vision retriever (Nemotron ColEmbed V2)~\cite{moreira2026nemotroncolembedv2topperforming}. For reproducibility and efficient evaluation, we precompute page embeddings offline and store them together with their page identifiers. This design makes the online retrieval cost independent of corpus size beyond the index lookup: at test time, LITTA only encodes query variants and scores a fixed candidate depth per variant before fusion.

\subsection{Multi-Query Retrieval and Fusion}

Given an original query, LITTA generates $Q-1$ expanded queries (query variants) and performs retrieval for each query independently using late interaction (MaxSim)~\cite{khattab2020colbert}. We then aggregate candidates from expanded-query retrieval using Reciprocal Rank Fusion (RRF)~\cite{cormack2009rrf}, which emphasizes pages that are consistently ranked highly across variants.

Let $r_i(p)$ denote the rank of page $p$ under the $i$-th query variant. The fused score is:
\begin{equation}
\mathrm{RRF}(p)=\sum_{i=0}^{Q-1}\frac{1}{k_{\mathrm{RRF}} + r_i(p)}.
\end{equation}
We use $k_{\mathrm{RRF}}{=}60$ by default and select the final top-$K$ pages after fusion.

Unless stated otherwise, we use:
\begin{itemize}
\item number of queries: $Q \in \{1,3,5\}$,
\item retrieval depth per query: a fixed top-$k$ candidate depth per query (kept constant across runs),
\item pooling and ranking: Reciprocal Rank Fusion with a fixed $k_{\mathrm{RRF}}$,
\item final output: top-$K$ pages after fusion.
\end{itemize}

%%%%%%%%%%%%%%%%%%%%%%%%%%%%%%%%%%%%%%%%%%%%%%%%%%%%%%%%%%%%%
\section{Algorithm and Mathematical Formulation}
\label{sec:appendix_algorithm}
%%%%%%%%%%%%%%%%%%%%%%%%%%%%%%%%%%%%%%%%%%%%%%%%%%%%%%%%%%%%%

We use ColBERT-style late interaction (MaxSim) for scoring. Let $T = \{\mathbf{t}_i\}_{i=1}^{N}$ be query token embeddings and $P = \{\mathbf{p}_j\}_{j=1}^{M}$ be page patch/token embeddings:
\begin{equation}
S(T,P) = \sum_{i=1}^{N} \max_{j \in \{1,\ldots,M\}} \left( \mathbf{t}_i^\top \mathbf{p}_j \right).
\end{equation}

For multi-query retrieval, we aggregate candidate rankings across expanded queries and select final evidence pages from the pooled list.

%%%%%%%%%%%%%%%%%%%%%%%%%%%%%%%%%%%%%%%%%%%%%%%%%%%%%%%%%%%%%
\section{Additional Analysis}
\label{sec:appendix_additional}
%%%%%%%%%%%%%%%%%%%%%%%%%%%%%%%%%%%%%%%%%%%%%%%%%%%%%%%%%%%%%

Across domains, increasing $Q$ generally improves retrieval quality by recovering complementary evidence pages that may be missed by the original query alone. In practice, $Q=3$ often provides a strong accuracy--cost trade-off, while $Q=5$ can further improve difficult domains (e.g., industrial manuals) at higher inference cost.

\looseness=-1
We also find that the benefit of multi-query retrieval correlates with terminology heterogeneity and the degree to which evidence is visually localized. When relevant information is concentrated in tables or schematics, small lexical changes (e.g., adding an acronym or a part name) can activate different regions under late interaction, and fusion helps consolidate these complementary signals into a single robust ranking.

\end{document}